\documentstyle[12pt,world_sci,epsf]{article}
\begin{document}

\title{{\bf MECHANICAL DESIGN OF THE CDF SVX~II SILICON VERTEX DETECTOR}}
\author{JOHN E. SKARHA\thanks{For the CDF SVX~II Group;
Currently Guest Scientist at Fermi National Accelerator Laboratory,
P.O. Box 500, M.S. 318,
Batavia, IL 60510, USA}\\
Department of Physics and Astronomy\\
The Johns Hopkins University\\
Baltimore, Maryland 21218, USA}
\vspace{0.3cm}

\maketitle
\setlength{\baselineskip}{2.6ex}

\begin{center}
\parbox{13.0cm}
{\begin{center} ABSTRACT \end{center}
{\small \hspace*{0.3cm} A next generation silicon vertex detector is planned
at CDF for the 1998 Tevatron collider run with the Main Injector.
The SVX~II silicon vertex detector will
allow high luminosity data-taking, enable online triggering of secondary
vertex production, and greatly increase the acceptance for heavy flavor physics
at CDF.  The design specifications, geometric layout, and early
mechanical prototyping work for this
detector are discussed.}}
\end{center}

\section{Introduction}
	The heavy flavor physics program of the Collider Detector at Fermilab
(CDF~\cite{CDF}) has greatly benefitted from the successful operation of the
CDF SVX silicon vertex detector~\cite{svx}.  Precise SVX position measurements
have been used in the tagging of $b$-jets to identify top quark candidate
events~\cite{top} and for lifetime measurements of $b$-hadrons~\cite{blife}.
The need for high resolution
vertex detection
will become even more
important with the high luminosity ($\sim 10^{32}$ cm$^{-2}$ sec$^{-1}$)
running
and shorter bunch spacing (132 or 395 ns) during the Main Injector era starting
in 1998.  The change in bunch spacing forces the replacement of the present
radiation-hard SVX$^\prime$ detector~\cite{prem}.  A second generation
silicon vertex
detector, the
SVX~II~\cite{svxii}, is currently in the design and early prototyping
stages and
will double the geometric coverage of the long interaction region
at the Tevatron and extend the tracking coverage in pseudo-rapidity.  The
use of double-sided detectors will provide 3D vertex reconstruction,
improving the background rejection in both top and $b$ physics analyses.
A Level 2 trigger processor, the Silicon Vertex
Tracker (SVT)~\cite{SVT}, will apply impact parameter cuts using the vertex
detector information for tracks found at Level 1 in the outer central tracking
chamber.  Such vertex based triggering is particularly important for high
statistics studies of $B$ decay modes such as $B^0 \rightarrow \pi^+\pi^-$.

\section {\bf General Description}
In order to reduce the channel occupancy and the input capacitance for the
readout electronics, the SVX~II detector will be arranged in three
identical barrel modules
mounted symmetrically with respect to the interaction point (Fig.~1).
\begin{figure}[thb]
\vspace{-1.0cm}
\parbox[b]{5.5in}{\epsfxsize=5.5in\epsfbox{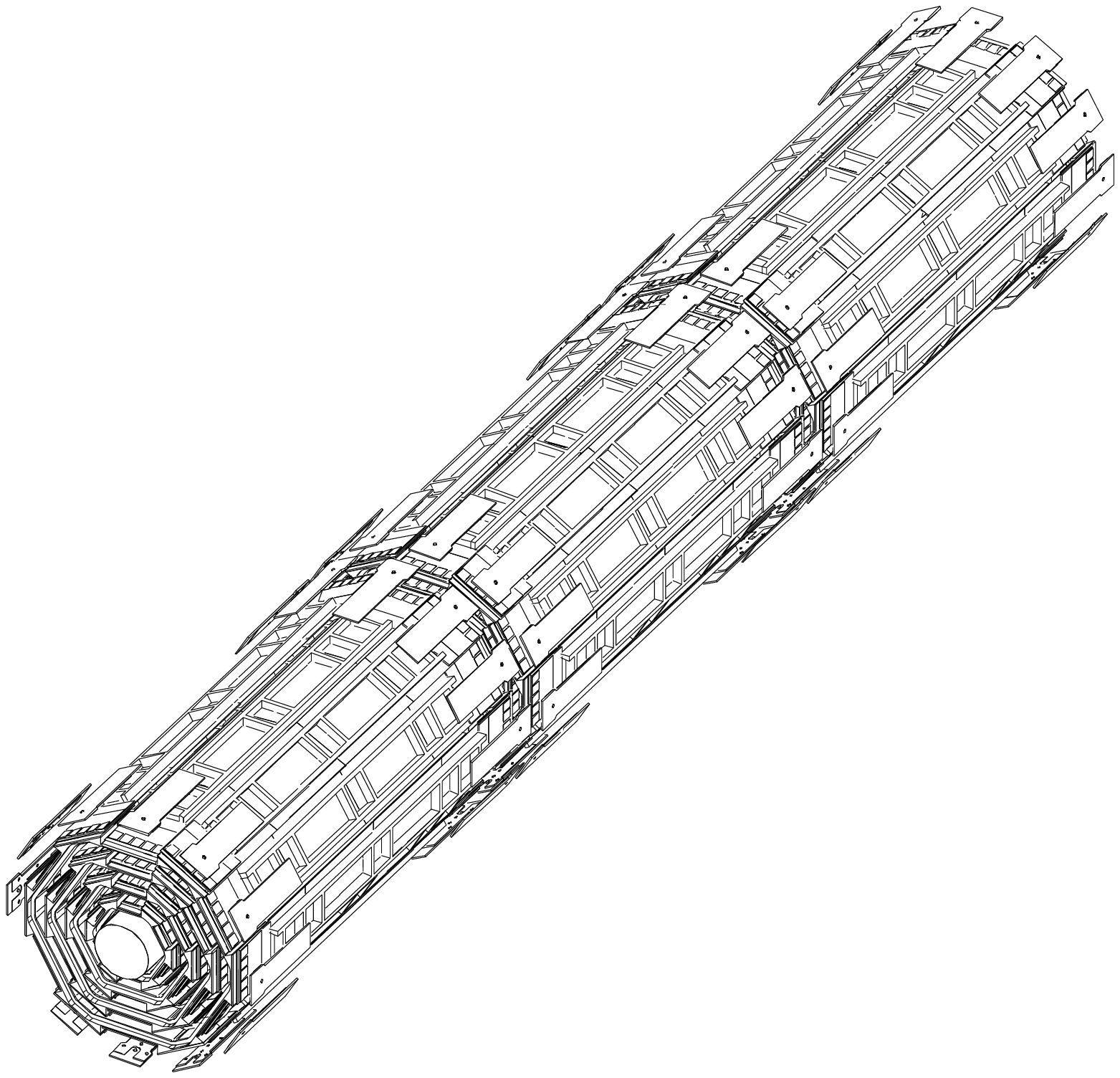}}%
\vspace{-1.0cm}
\begin{center}
{\small Fig. 1: The three barrels of the CDF SVX~II silicon vertex detector.}
\end{center}
\label{barrel3}
\end{figure}
Each barrel will
be 32 cm in length, with four layers of silicon (numbered 0 to 3) grouped into
ladders and arranged in
12 wedges in azimuth.
Each ladder will consist of four
silicon detectors mounted together into a single mechanical unit and
wirebonded electrically in pairs which are read out at each end of the ladder.
In order to reduce the dead space between barrels, the kapton readout hybrids
will be attached to a beryllium substrates and mounted on the surface of the
outer detectors of a ladder.  The kapton/beryllium combination should have
good cooling performance and lower mass than standard thick film hybrids.
The ladders themselves are mounted between two
precision machined beryllium bulkheads which have a staggered radii geometry
and
allow for overlap between neighboring ladders.

Figure~2 shows the
geometry of the SVX~II bulkhead.  In contrast to SVX, the SVX~II bulkhead has
an integrated cooling channel within each mounting ring.  This is
accomplished by
4 separate cooling rings which are glued to the mounting rings to form the
cooling channel for each layer.  There is 1 input and output coolant
connection for
each layer.
Such a design improves the cooling performance and minimizes the
overall bulkhead plus cooling system mass.
\begin{figure}[thb]
\vspace{-1.0cm}
\parbox[b]{5.5in}{\epsfxsize=5.5in\epsfbox{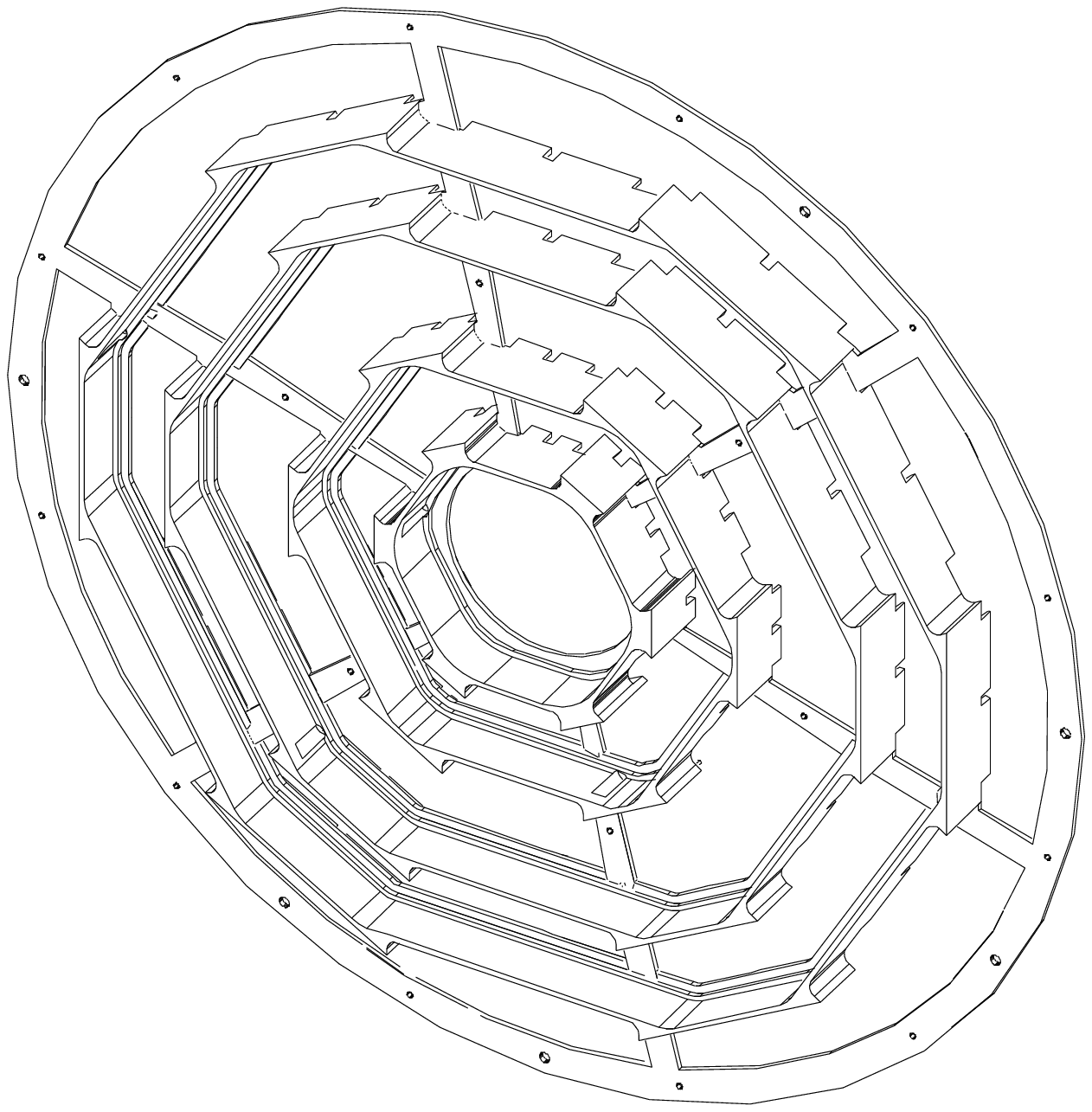}}%
\vspace{-1.0cm}
\begin{center}
{\small Fig. 2: A solid model of the SVX~II bulkhead.}
\end{center}
\label{bulk}
\end{figure}

The SVX~II ladder design (Fig.~3) has evolved from that used for SVX and uses a
structure formed out of boron-carbon fiber and Rohacell foam
to couple 2 two-detector half-ladders together to
form a complete 4 detector ladder.  Significant engineering effort has gone
into designing the ladder support to have the same thermal expansion
coefficient
as silicon in order to minimize stresses in the silicon during temperature
changes.
The ladder support structure should maintain detector-to-detector
alignment to within $\pm$5 microns in the $r-\phi$ direction.
Since the double-sided detectors~\cite{seidel} in the present SVX~II design
have an orthogonal
stereo angle to optimize the $r-z$ vertex resolution, a tight tolerance on
the radial uncertainty of the detectors is also required.  This is because for
tracks at large incident angles, there is a strong coupling between the radial
and $z$ position uncertainties.  The $z$ position uncertainty in the placement
of the detectors during the ladder
construction process should be $\sim \pm$10 microns.

\begin{figure}[thb]
\vspace{-1.0cm}
\parbox[b]{5.5in}{\epsfxsize=5.5in\epsfbox{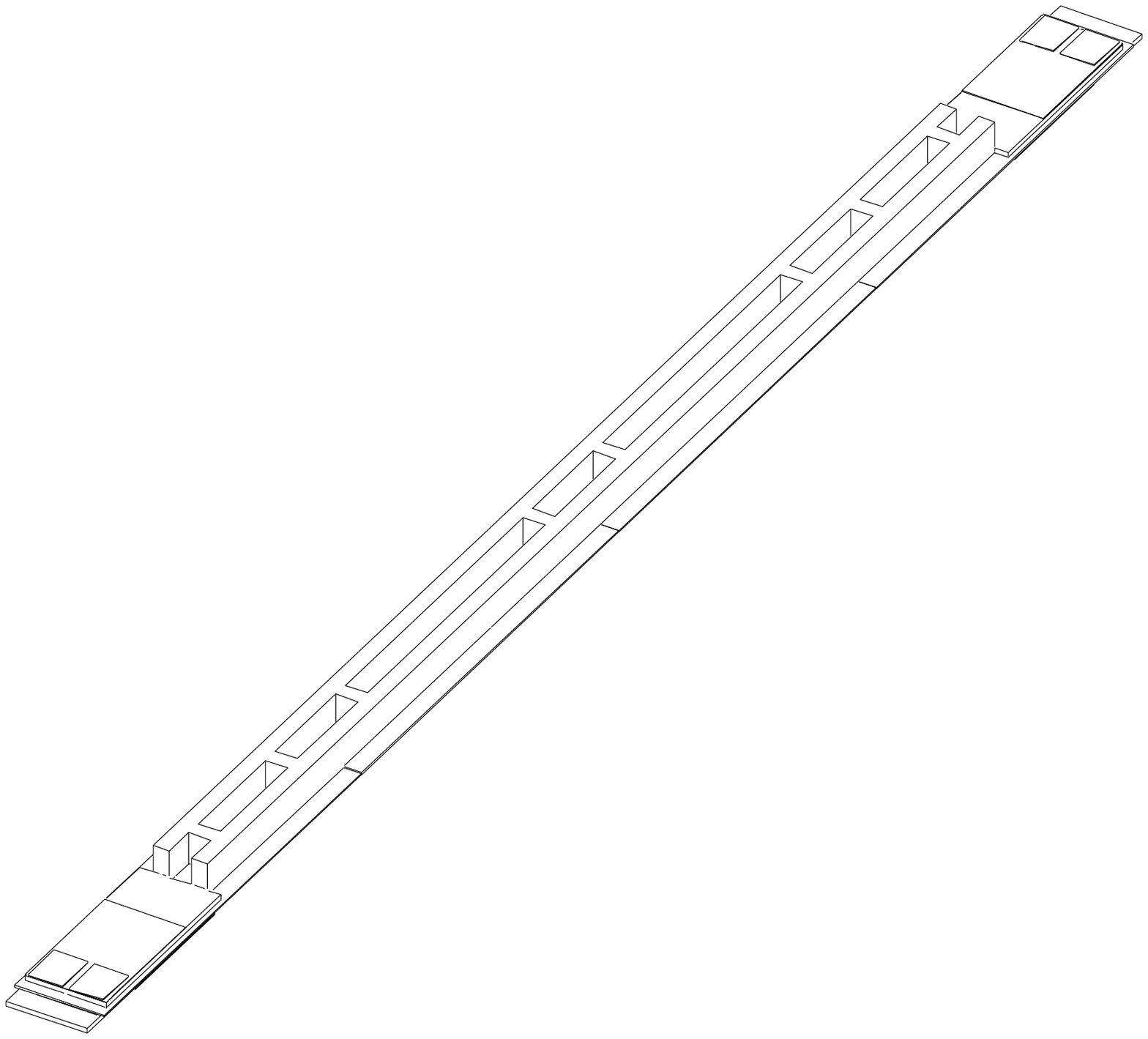}}%
\vspace{-1.0cm}
\begin{center}
{\small Fig. 3: Solid model of the SVX~II layer 0 ladder.}
\end{center}
\label{ladder}
\end{figure}
Due to the double-sided readout and having readout at both ends of the
ladder, the power generated by the readout chips is substantial
($\sim 300 - 500$ milliwatts/chip).
In addition, there is the power generated by the port card
readout electronics~\cite{gold} (up to a maximum of 6~watts per port card).
Because of
an estimated 1.4 kilowatts total power produced by the three barrel
combination,
lower temperature alcohol-based coolants delivered near $0 ^{\circ}$C are being
considered as the best way of cooling the silicon to possibly $10 ^{\circ}$C
in order to maintain good signal-to-noise after radiation damage.  Low
temperture
dry gas will also have to be delivered to the closed detector volume in order
to cool the silicon away from the bulkhead.

In order to achieve the 100~$\mu$radian alignment of the detector $r-\phi$
strips to the beam required by the SVT trigger to avoid high rates from fake
impact parameter triggers, the three barrels will be mounted in an external
space frame.  Early designs of the space frame consist of a structurally
strong, low mass web of carbon fiber strips molded into a half cylinder.
The space frame is estimated to weigh less than 0.5 kg and
have a deflection of $< 20 \mu$m when fully loaded.  The space frame will also
provide mounting support for the portcards, cables, and cooling pipes.

Radiation damage is certainly expected to affect the performance of the SVX~II
detector over the lifetime of its operation.  The silicon layer closest
to the beam is most affected since the radiation dose has been observed
experimentally to be proportional to $r^{-1.7}$.  The silicon strip
detectors are expected to
go through type inversion during colliding beam operation
and layer 0 will probably need to be replaced after $\sim 1.5$ Mrad exposure.

\section{Comparison with SVX}
Table~\ref{design_comp} shows a comparison
between the SVX and SVX~II detector parameters.  We see that the most
dramatic difference is the doubling of the length has tripled the number of
$r-\phi$ readout channels (due to the shorter electrical length) and that the
addition of the $r-z$ readout nearly doubles again the number of channels.

\section{Conclusions}
We have described the mechanical design of the CDF SVX~II vertex detector
for Run~II.
This  detector has double the barrel length over the present SVX$^{\prime}$,
uses double-sided silicon detectors, and allows for the possibility of
triggering on high impact parameter tracks.
The increased acceptance, efficiency, and background rejection
should significantly improve the CDF top and bottom physics capabilities during
the Fermilab Main Injector high luminosity running!
\begin{table}[htb]
\begin{center}
{\small Table 1: Comparison of the SVX and SVX~II detector geometries.}
\begin{tabular}{|c|c|c|}\hline
Detector Parameter          & SVX    & SVX~II \\ \hline
Readout coordinates         & $r$-$\Phi$ & $r$-$\Phi$;$r$-$z$ \\
Number of barrels           &   2      &   3  \\
Number of layers per barrel &   4      &   4  \\
Number of wedges per barrel &  12      &  12  \\
Ladder length               &  25.5 cm  &  32.0 cm \\
Combined barrel length      & 51.0 cm & 96.0 cm \\
Layer geometry              & 3$^{\circ}$ tilt & staggered radii \\
Radius innermost layer      & 3.0 cm & 2.4 cm \\
Radius outermost layer        & 7.8 cm & 8.7 cm \\
$r$-$\phi$ readout pitch (4 layers) & 60;60;60;55 $\mu$m & 60;62;58;60 $\mu$m
\\
$r$-$z$ readout pitch (4 layers) &     -              & 149;132;99;149 $\mu$m
\\
Length of readout channel ($r$-$\phi$) & 25.5 cm  & 16.0 cm \\
$r$-$\phi$ readout chips/ladder (4 layers) & 2;3;5;6 & 4;6;10;12 \\
$r$-$z$ readout chips/ladder (4 layers) & - & 4;6;8;8 \\
$r$-$\phi$ readout channels & 46,080 & 147,456 \\
$r$-$z$ readout channels   &   -       & 119,808 \\
Total number of channels    & 46,080    & 267,264 \\
Total number of readout chips & 360  & 2088 \\
Total number of detectors & 288     & 576 \\
Total number of ladders     &  96       &  144   \\
Silicon area (m$^2$)          & 0.68   & 1.5 \\
Diode length (miles)    & 7.3   & 17.5 \\
\hline
\end{tabular}
\end{center}
\label{design_comp}
\end{table}

\end{document}